\documentclass[a4paper,fleqn]{cas-sc}

\usepackage[numbers]{natbib}
\usepackage{amsmath}
\usepackage{graphicx}
\usepackage{array}
\usepackage{tabularx}
\usepackage{enumitem}
\usepackage{xcolor}
\usepackage{csquotes}
\usepackage[gen]{eurosym}
\usepackage{setspace}

\usepackage{longtable}
\usepackage{adjustbox}

\DeclareMathOperator*{\argmin}{argmin}
\graphicspath{{../graphics/}}

\def\tsc#1{\csdef{#1}{\textsc{\lowercase{#1}}\xspace}}
\tsc{WGM}
\tsc{QE}

\begin{document}
\let\WriteBookmarks\relax
\def\floatpagepagefraction{1}
\def\textpagefraction{.001}

\shorttitle{}    

\shortauthors{}  

\title [mode = title]{On Optimal Battery Sizing for Electric Vehicles}  



%

\author[1]{Felix Wieberneit}
\cormark[1]
\ead{fw1520@ic.ac.uk}
\credit{Investigation, Software, Visualization, Writing - original draft}

\author[2]{Emanuele Crisostomi}
\ead{emanuele.crisostomi@gmail.com
}
\credit{Conceptualization, Writing - review and editing}

\author[1,3]{Anthony Quinn}
\ead{a.quinn@imperial.ac.uk}
\credit{Conceptualization, Writing - review and editing}

\author[1]{Homayoun Hamedmoghadam}
\ead{h.hamed@imperial.ac.uk}
\credit{Conceptualization, Writing - review and editing}

\author[1]{Pietro Ferraro}
\ead{p.ferraro@imperial.ac.uk}
\credit{Conceptualization}

\author[1]{Robert Shorten}
\ead{r.shorten@imperial.ac.uk}
\credit{Conceptualization, Supervision}

\affiliation[1]{organization={Dyson School of Design Engineering, Imperial College London},
              country={United Kingdom}}

\affiliation[2]{organization={Department of Energy, Systems, Territory and Constructions Engineering, University of Pisa},
              country={Italy}}

\affiliation[3]{organization={Electronic and Electrical Engineering, Trinity College Dublin},
                            country={Ireland}}

\cortext[1]{Corresponding author}

\fntext[1]{}


\begin{abstract}
In this paper, we introduce a quantitative framework to optimize electric vehicle (EV) battery capacities, considering two criteria:  upfront vehicle cost and charging inconvenience cost.
For this purpose, we (1) develop a comprehensive model for charging inconvenience costs, incorporating both charging time and detours, improving on existing studies, (2) show, through extensive simulations and analytical models, how charging inconvenience cost is affected by different battery capacity and charging infrastructure configurations, (3) introduce an optimisation framework to determine optimal battery capacities based on charging inconvenience and vehicle cost, and (4) show that optimal battery capacities can be influenced by strategic investments in charging infrastructure and tax/incentive policies.
The proposed framework provides actionable insights into the sustainable design of  EV systems, supporting the development of cost-effective and convenient electric mobility solutions.
\end{abstract}


\begin{highlights}
\item We optimize battery capacities based on vehicle and charging inconvenience costs.
\item Traffic simulations quantify the impact of battery and infrastructure on inconvenience costs.
\item Larger battery capacities reduce inconvenience costs, but the incremental benefit diminishes.
\item Optimal battery capacities---which minimize vehicle and inconvenience costs---are smaller than expected.
\item Investments in infrastructure can drive  optimal battery capacities towards smaller, more sustainable values.
\end{highlights}

\begin{keywords}
 Electric vehicles (EVs)\sep Battery capacity \sep Charging infrastructure\sep Charging inconvenience \sep Optimization
\end{keywords}

\maketitle

\doublespacing
\section{Introduction}
\label{sec:intro}

Electric vehicles (EVs) are integral to the global effort to reduce greenhouse gas emissions and combat climate change \cite{iea2024, UNEP2020}. Their widespread adoption is critical for achieving sustainability goals and enhancing the quality of lives in both urban and rural areas \cite{UNEP2020}. 

In this paper, we introduce a quantitative framework to determine optimal EV battery capacities. This framework seeks to minimize the private costs incurred by EV drivers based on two criteria: upfront purchase cost of vehicles, and charging inconvenience. Both of these criteria are dependent on the battery capacity of the EV, and are critical for broad adoption of EVs, as discussed next.

The persistently \textbf{high upfront costs} of EVs are strongly influenced by the cost of battery packs. While battery costs have declined significantly since 2007 \cite{nykvist2015, bloombergnef2023, frith2023}, the upfront cost of EVs is still higher than that of comparable internal combustion engine vehicles (ICEVs)~\cite{nykvist2015}. This economic barrier continues to deter potential buyers. \cite{Li2017, krishna2021}.

Similarly, \textbf{charging inconvenience} continues to affect consumer acceptance of EVs. Despite the advances in battery technology \cite{bloombergnef2023, frith2023}, EVs still face challenges related to charging times and frequency, which can make their use less convenient in comparison to  ICEVs \cite{dixon2020}. This problem is particularly relevant for drivers without private charging access, such as those living in dense urban centres without private driveways.

 Automakers have increased EV battery capacities  in recent years, with the aim of alleviating charging inconvenience \cite{iea2024}. This trend, often referred to as the {\em EV obesity problem\/}  \cite{economist2023}, results in  EVs that are considerably more expensive---and heavier---than their ICEV counterparts. 

Clearly, an alternative to increasing battery capacities is to invest in a dense and high-powered charging infrastructure. The  complementary roles that batteries and charging infrastructure play in the design of driver-friendly EV systems are recognised in the literature. However, there remains an unresolved debate.
Some authors argue in favour of larger batteries, stating that these reduce the need for extensive charging networks \cite{wenig2019}, while others contend that improved charging infrastructure allows for smaller, more efficient batteries \cite{jones2023a, neubauer2014}.

A  full understanding of this trade-off has not yet been reached. Existing studies neglect drivers who rely on public charging infrastructure \cite{wenig2019}, or quantify charging inconvenience with simplified models that do not account for the battery-infrastructure trade-off \cite{dixon2020}.
What is missing is a quantitative framework that integrates these complementary perspectives on battery size and charging infrastructure.
In this paper, we introduce such a framework, focussing particularly on public charging infrastructure and on identifying an optimal battery capacity that minimizes both the upfront vehicle cost and the charging inconvenience in a given charging environment.
Our key contributions are:

\begin{enumerate}
    \item \textbf{Charging Inconvenience Indicator:} We introduce a model that quantifies the private inconvenience (for the driver) of charging their EV. We do this more comprehensively than in previous models \cite{dixon2020}, by considering both the time spent charging, as well as the time for charging detours. Using this inconvenience indicator, we explore the impact of different battery and charging infrastructure configurations on charging inconvenience using analytical models and extensive traffic micro-simulations.

    \item \textbf{Diminishing Returns from Larger Batteries:} Both analytical models and traffic simulations clearly demonstrate that---beyond a certain point---increasing battery capacity yields minimal reductions in charging inconvenience. This suggests that bigger batteries  may not be  effective in reducing charging inconvenience, and that investment in  charging infrastructure (e.g.\ a higher density or faster charging infrastructure) may be a more effective way to increase the convenience of EVs.

    \item \textbf{Optimization Framework for Identifying Optimal Battery Capacities:} We develop a quantitative framework to determine  optimal battery capacity, in the sense that it  jointly minimizes charging inconvenience and vehicle cost. Notably, this framework captures the effect of battery capacity, charging infrastructure density, power, and charge-point competition among EV drivers. Our findings suggest that the optimal capacities in many urban settings are significantly smaller than those provided by many current EV models on the European market.

    \item \textbf{Influencing the Optimal Battery Capacity:} We show that the optimal battery capacity is not fixed but can be strategically influenced through interventions such as expanding fast-charging infrastructure and implementing taxation or incentive schemes. This highlights the role of policy-makers and urban planners in shaping sustainable EV systems.
\end{enumerate}

By considering the tradeoffs between batteries and infrastructure, and identifying locally optimal battery capacities, cities can design targeted infrastructure investments and tax/ incentive policies, making the use of EVs more convenient and affordable. Our findings offer valuable insights for policymakers, urban planners, and manufacturers aiming to foster sustainable EV adoption in urban settings.

\subsection{Structure of the Paper}
In Section~\ref{sec:B^o}, we formalise the optimisation problem, deducing an optimal battery capacity, $B^o$, that jointly minimizes charging inconvenience and purchase cost of EVs, depending on individual travel needs and the charging environment. Furthermore, we detail how we estimate purchase cost and charging inconvenience, introducing both analytical and simulation approaches. In Section \ref{sec:scenarios}, we describe four e-mobility scenarios that we have designed to isolate the effect of battery and charging infrastructure variables on charging inconvenience. In Section \ref{sec:central_paris}, we present a realistic case study of central Paris, to validate our results. Based on this case study, we also apply our optimization framework to a simulated population of Parisians and compare the results to the European EV market. In Section \ref{sec:influencing_b^o}, we explore how the optimal battery capacity, $B^o$, can be influenced, detailing how investments in charging infrastructure density, as well as targeted tax-incentive programs may shift $B^o$ towards smaller, more sustainable batteries. Finally, in Section~\ref{sec:conc}, we discuss the main findings of the study, and we conclude the manuscript by outlining future directions which may  extend the framework developed here.

\begin{table}[htbp!] 
  \begin{minipage}{\linewidth}
    \centering
    \caption{Notation used in the analysis of optimal EV battery capacities}
    \label{tab:new_symbols}
    \begin{tabular*}{\tblwidth}{@{}LL@{}}
        \hline
        Symbol & Description \\
        \hline
        $B$ & Battery capacity (kWh) \\
        $B^o$ & Optimal battery capacity (kWh) \\
        $\rho$ & Charging infrastructure density (number of charging stations per km$^2$) \\
        $\xi$ & Charging infrastructure utilization rate \\
        $c_p(B)$ & Annualised purchase cost of EV after subsidies as a function of battery capacity, $B$ (Euro) \\
        $c_e(B, d, \rho, \xi)$ & Inconvenience cost (Euro) associated with charging an EV, depending on $B$, $d$, $\rho$, and $\xi$ \\
        $p$ & Value of tax/incentive policies affecting the effective purchase price (Euro) \\
        $c_{p,base}(B)$ & Base price of the EV (before subsidies) as a function of battery capacity $B$ (Euro) \\
        $\beta$ & Residual value of the EV after its lifetime, as a fraction of initial purchase cost \\
        $T$ & Expected lifetime of the EV (years) \\
        $\tau_e(B, d, \rho, \xi)$ & Total inconvenience time incurred due to charging (hours) \\
        $\mu$ & 'Time is money' factor for time in urban transportation (Euro/hour) \\
        $d$ & Annual driving distance without charging detours (km) \\
        $\tau_{p,o}(d)$ & Total time to charge energy for the originally intended travel distance $d$ (hours) \\
        $\tau_{p,s}(B, d, \rho, \xi)$ & Additional charging time to replenish energy consumed during charging detours (hours) \\
        $\tau_s(B, d, \rho, \xi)$ & Time spent traveling to and locating available charging stations (hours) \\
        $\eta$ & Electricity consumption rate of the EV (kWh/km) \\
        $P$ & Power output of charging stations (kW) \\
        $v$ & Average travel speed while searching for charging stations (km/h) \\
        $d_s(B, d, \rho, \xi)$ & Total additional distance traveled in search of charging stations (km) \\
        $n_c(B)$ & Number of charging events required for a total distance $d + d_s(B, d, \rho, \xi)$ \\
        $\sigma$ & Average fraction of battery capacity charged during charging events \\
        $d_{s,k}(\rho, \xi)$ & Expected detour distance to find an available charging station during the $k$-th charging event (km) \\
        \hline
    \end{tabular*}
    
  \end{minipage}
\end{table}
\section{Optimization Framework for EV Battery Capacities}
\label{sec:B^o}

The primary objective of this paper is to determine the optimal battery capacity, $B^o$, of  an EV. This should take account of the annualised purchase cost of the EV, $c_p$, along with the annual charging inconvenience cost, $c_e$, both of which depend on battery capacity, $B$. The optimization problem is formulated as follows:
\begin{equation}
    B^o = \argmin_{B>0} [ c_p(B) + c_e(B, d, \rho, \xi) ]
    \label{eq:optimal_battery}.
\end{equation}
Here: 
\begin{itemize} 
\item $d$ is the annual distance driven by the EV;
\item $\rho$ is the density of charging stations (CSs) (influencing the distance that needs to be covered to recharge an EV);
\item $\xi$ is the utilization rate of the charging infrastructure (indicating the likelihood of stations being occupied). 
\end{itemize}

Intuitively, as battery capacity $B$ increases, $c_p$ generally increases due to the higher cost of larger batteries. Conversely, all other variables being equal, $c_e$ tends to decrease with larger battery capacities, as drivers need to charge less frequently. (\ref{eq:optimal_battery}) seeks an optimal trade-off  between these opposing trends, yielding a battery capacity---as a function of $d$, $\rho$ and $\xi$---that minimizes the total cost to the driver.

In Sections~\ref{sec:c_p} and \ref{sec:c_e}, respectively, we detail the components of the optimisation problem in \eqref{eq:optimal_battery}, by describing how we estimate the annualised purchase cost after subsidies $c_p$ and the charging inconvenience cost $c_e$.

\subsection{EV Purchase Cost after Subsidies, $c_p$}
\label{sec:c_p}

To estimate the purchase cost of an EV, we note the following: 
\begin{enumerate}
    \item Operational costs (e.g., electricity consumption, maintenance) are not included in this analysis;
    \item  The base price $c_{p,base}(B)$ of EVs depends solely on their battery capacity;
    \item The vehicle purchase cost is spread evenly across the  expected lifetime, $T$, of the EV (typically $T = 10$ years), after which it is sold at a proportion $\beta$ of the initial purchase cost (we assume $\beta=0.25)$.
\end{enumerate}

The annualised purchase cost $c_p(B)$ after taxes/incentives is then calculated as
\begin{equation}
    c_{p}(B) = \frac{(1-\beta)\cdot(c_{p,base}(B)+p)}{T},
    \label{eq:ev_purchase_cost}
\end{equation}
where $p$ (Euro) denotes the total (signed) value of the potential tax or incentive policies that may affect the net purchase price, as explained in the last paragraph of this section.

We estimate the base price $c_{p,base}(B)$, using regression analysis on advertised EV prices and their corresponding battery capacities in the European market. Data from 309 EV models are obtained from a reputable EV database \cite{ev-database.org}. These data are filtered to remove outliers based on the interquartile range (IQR) method ~\cite{barbato2011}. Specifically, data points with prices below $Q_1 - 1.5 \times IQR$ or above $Q_3 + 1.5 \times IQR$ (where $Q_{\cdot}$ denote the empirical quartiles for the data) are identified as outliers and excluded from the analysis. As a result, the data set is reduced from 309 to 277 data points.

Both linear and polynomial regressions are performed, with the polynomial regression yielding a significantly better fit (lower residual sum of squares). For the polynomial regression, a second-order polynomial is selected without further optimization of the order, as higher-degree polynomials appear not to provide significant improvements in fit. The models are trained on the pre-processed data set containing 277 points. The selected second order polynomial function for further analysis is: 
\begin{equation}
    c_{p,base}(B) = 5.113B^2 + 84.871B + 26316.599.
    \label{eq:purchase_cost_regression}
\end{equation}

\begin{figure}
    \centering
    \includegraphics[width=0.6\textwidth]{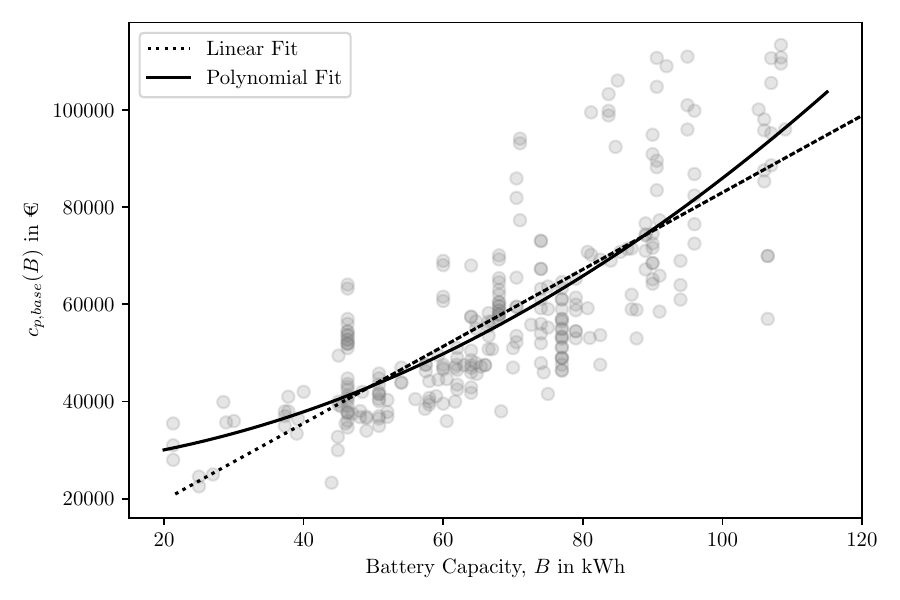}
    \caption{Regression analysis of $c_{p,base}$ as a function of $B$.}
    \label{fig:purchase_cost_regression}
\end{figure}

Figure \ref{fig:purchase_cost_regression} illustrates the individual data points and the linear and second order polynomial fits, showing the relationship between battery capacity and base price for the EV models considered. The total cash amount of any  tax ($p>0$) or incentive ($p<0$) policies  is then added to this base price \eqref{eq:purchase_cost_regression}, yielding the annualised cost of EV purchase after subsidies, $c_p(B)$, via \eqref{eq:ev_purchase_cost}. This provides the first term in the objective  \eqref{eq:optimal_battery}. 

\subsection{Charging Inconvenience Cost $c_e$}
\label{sec:c_e}

While prior studies related to EV inconvenience have predominantly focused on optimizing the placement and sizing of charging infrastructure to minimize the inconvenience experienced by EV drivers \cite{xu2022,davidov2020,levinson2018}, our research introduces a crucial new dimension by directly linking battery capacity to driver inconvenience. Previous work, such as that by Dixon et al. \cite{dixon2020}, has quantified the inconvenience of EV charging as a time-penalty, based solely on the time required to physically charge the vehicle {\em en route}. The authors use this term to refer to the specific charging events when EVs have to take a detour from their driving patterns to stop for charging at a (usually) fast charging point, before proceeding to their next destination. We extend the definition of charging inconvenience to include not only the time spent physically charging but also the additional time incurred due to detours for charging and potential delays in finding available CSs. This more accurately represents the overheads faced by EV drivers.

Our charging inconvenience time indicator, $\tau_e(B, d, \rho, \xi)$ quantifies the total annual time cost associated with charging, based on the total distance driven annually $d$; the battery capacity $B$; the density of CSs $\rho$; and the utilization rate of CSs $\xi$. The derivation of $\tau_e$ is described in Section \ref{sec:inconvenience_time}. In order to combine this with $c_p(B)$ into a Euro-valued objective \eqref{eq:optimal_battery}, 
 $\tau_e$ is converted into a monetary inconvenience cost (again in Euro), $c_e$, using a standard time-is-money (i.e.\ valuation) factor $\mu$. This conversion is detailed in Section \ref{sec:conversion}

\subsubsection{Analytical Estimation of Charging Inconvenience Time $\tau_e$}
\label{sec:inconvenience_time}
To derive an analytical expression for $\tau_e(B, d, \rho, \xi)$, we make the following simplifying assumptions (these assumptions will be dropped in Section \ref{sec:simulation}, where $\tau_e$ is computed through data from traffic simulations):

\begin{enumerate}
    \item The electricity consumption rate $\eta$ of the EVs remains constant regardless of $B$. Typical values range from 0.15 to 0.25 kWh/km;
    \item The expected proportion of the battery capacity charged at each charging event $\sigma$, is the same across all battery capacities. A typical value is around 0.7;
    \item The charge rate of CSs ($P$) and the average travel speed during detours to CSs ($v$) are taken as fixed values, for example $P = 20$ kW and $v = 15$ km/h. 
\end{enumerate}
Under these assumptions, the total inconvenience time $\tau_{e}(B, d, \rho, \xi)$ can be expressed as the sum of three terms,
\begin{equation}
    \tau_e(B, d, \rho, \xi)=\tau_{p,o}(d)+\tau_{p,s}(B, d, \rho, \xi) + \tau_{s}(B, d, \rho, \xi)
    \label{eq:inconvenience2},
\end{equation}
where 
\begin{itemize}
    \item $\tau_{p,o}(d)=\frac{\eta d}{P}$ is the total charging time required   for the planned travel distance $d$; this term is independent of $B$;
    \item $\tau_{p,s}(B, d, \rho, \xi)= \frac{\eta d_s(B, d, \rho, \xi)}{P}$ is the additional charging time due to the extra energy consumed during charging detours;
    \item $\tau_{s}(B, d, \rho, \xi)= \frac{d_s(B, d, \rho, \xi)}{v}$ is the time spent locating, and travelling to,  available CSs.
\end{itemize}

The total additional distance travelled for charging detours, $d_s(B, d, \rho, \xi)$, depends on the number of charging events and the average detour distance per charging event. The number of charging events $n_c(B)$ is:

\begin{equation}
   n_c(B) = \frac{\eta (d+d_s(B, d, \rho, \xi)) }{\sigma B}
    \label{eq:Number_Charging_Events},
\end{equation}
which accounts for the energy needed for both the planned travel and detours.

The total detour distance $d_s(B, d, \rho, \xi)$ is:
\begin{equation} d_s(B, d, \rho, \xi) = n_c(B,d) \cdot d_{s,k}(\rho, \xi), 
\label{eq:total_detour_distance} 
\end{equation} 

where $d_{s,k}(\rho, \xi)$ is the expected detour distance to find an available CS.
Solving for $d_s(B,d,\rho,\xi)$, we obtain:
\begin{equation}
    d_s(B, d, \rho, \xi) = \frac{\eta d \cdot d_{s,k}(\rho, \xi)}{\sigma B - \eta d_{s,k}(\rho, \xi)}
    \label{eq:d_s2},
\end{equation}
assuming $\sigma B > \eta d_{s,k}(\rho, \xi)$ to ensure a positive denominator.

To compute $ d_{s,k} $, we assume that---depending on the utilization of the charging infrastructure---multiple attempts may be necessary to find an available CS. The distance $ d_{s,k} $ is then given by:
\begin{equation} 
d_{s,k}(\rho, \xi) = \frac{1}{(1 - \xi)\cdot 2\sqrt{\rho}}, 
\label{eq:d_sk} 
\end{equation}
where
\begin{itemize}
    \item $ \frac{1}{1 - \xi} $ is the expected number of attempts required to find an available CS, derived from the geometric distribution ~\cite{ross2010};
    \item $\frac{1}{2\sqrt{\rho}}$ is the expected Euclidean distance to the nearest CS, assuming a homogeneous Poisson process for station locations~\cite{baccelli2010};
    \item $\xi = \frac{n_{\text{EV}} \cdot d \cdot \eta}{n_{\text{CS}} \cdot 8760 \cdot P}$ is the utilization factor, defined as the ratio of the demand for charging to the supply of charging capacity. The factor, $8760$ (hours), is required if we consider a period of observation of one year.
\end{itemize}

This equation demonstrates that as the utilization factor $ \xi $ grows (toward 1), the expected distance $ d_{s,k} $ increases, reflecting the increased difficulty of finding an available CS.

\begin{figure}[htbp!]
    \begin{minipage}{\textwidth}
        \centering
        \includegraphics[width=1\textwidth]{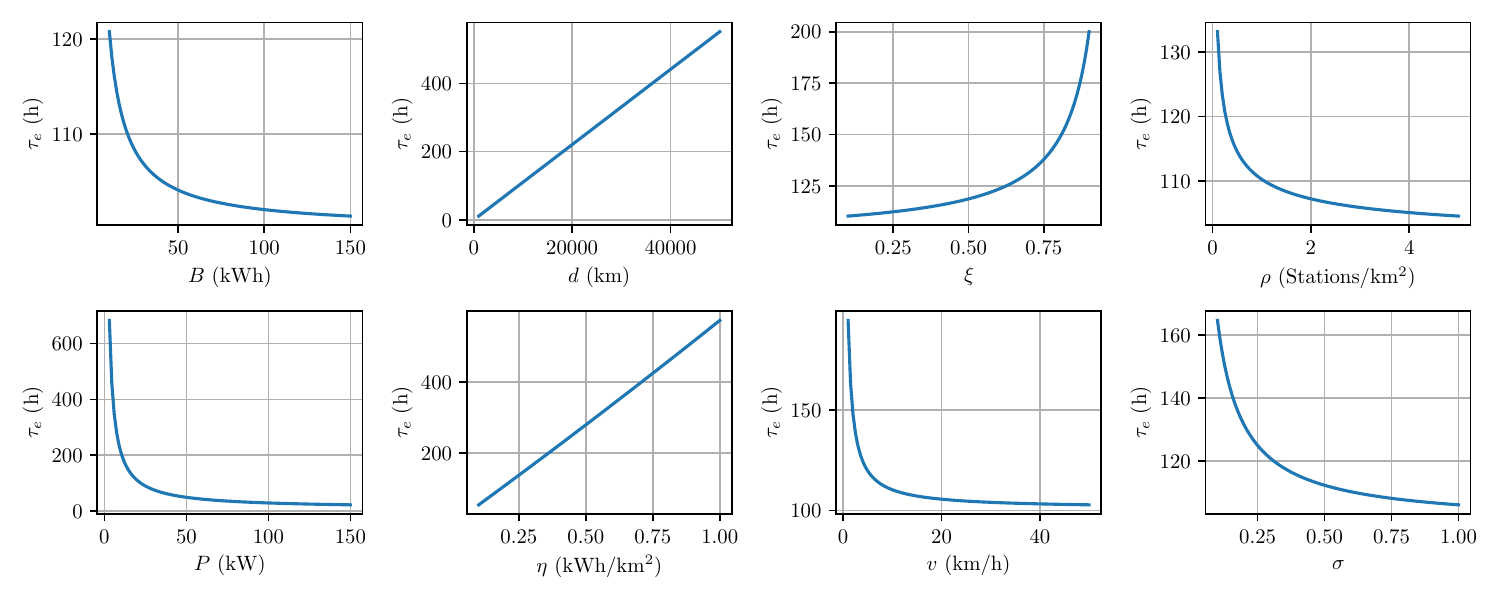}
        \caption{Sensitivity of charging inconvenience time $\tau_e$ with respect to its parameters.}
        \label{fig:tau_e_sensitivity}
    \end{minipage}
\end{figure}

By substitution into \eqref{eq:inconvenience2}, we obtain the total inconvenience time as
\begin{equation}
    \tau_e(B,d,\rho,\xi) = \frac{d \eta \left(0.5 P + v \cdot B \sqrt{\rho} \cdot \sigma (1 - \xi)\right)}{P v \left(B \sqrt{\rho} \cdot \sigma (1 - \xi) - 0.5 \eta\right)}
    \label{eq:expanded_inconvenience_time}.
\end{equation}
Hence, the inconvenience time $\tau_e(B, d, \rho, \xi)$ depends on a number of variables related to the EV itself ($B$, $\eta$), the behaviour of the driver ($\sigma$, $d$, $v$), as well as the charging infrastructure ($\rho$, $\xi$, $P$) and the urban environment ($v$). It can be minimized by strategically optimizing any of these  variables.
The sensitivity of \(\tau_e\) with respect to these variables is illustrated in figure \ref{fig:tau_e_sensitivity}, where we present a series of plots, each representing the effect of a single variable while keeping the others constant. The values for the constants are chosen to reflect a realistic scenario for EV operation, and the ranges for the variables of interest (listed in Table \ref{tab:variable_ranges}) are selected to account for the variability in these  values encountered in practice.

The analysis supports the following reasonable findings:
\begin{itemize} 
\item Increasing $B$ decreases $\tau_e$, but with diminishing returns beyond a certain point;
\item Higher CS density $\rho$, lower utilization rate, $\xi$ and higher charging power, $P$ also reduce $\tau_e$;
\item Improvements in vehicle efficiency $\eta$ or reductions in annual distance $d$ linearly decrease $\tau_e$.

\end{itemize}

\begin{table}[h!]
\caption{Constant values and ranges used to illustrate the sensitivity of $\tau_e$ in Figure \ref{fig:tau_e_sensitivity}.}
\label{tab:variable_ranges}
\begin{tabular*}{\tblwidth}{@{}LLL@{}}
\toprule
\textbf{Variable} & \textbf{Constant} & \textbf{Range} \\
\midrule
Efficiency \(\eta\) (kWh/km) & 0.2 & [0.1, 1] \\
Vehicle speed \(v\) (km/h) & 10 & [1, 50] \\
Charging factor \(\sigma\) & 0.6 & [0.1, 1] \\
Distance \(d\) (km) & 10,000 & [1,000, 50,000] \\
Battery capacity \(B\) (kWh) & 20 & [5, 150] \\
Utilization rate \(\xi\) & 0.1 & [0.1, 0.9] \\
CS density \(\rho\) (stations/km\(^2\)) & 1 & [0.1, 5] \\
Charging power \(P\) (kW) & 50 & [3, 150] \\
\bottomrule
\end{tabular*}

\end{table}

\subsubsection{Conversion of the Charging Inconvenience Time $\tau_e$ to Inconvenience Cost $c_e$}
\label{sec:conversion}

The time $\tau_e$ needs to be converted to a Euro-valued inconvenience cost, $c_e$ so that it is compatible with the objective in \eqref{eq:optimal_battery}. In this study, we consider the case of Paris and rely on valuation factors published by the French government for the economic evaluation of public investments \cite{quinet2013, meunier2020}. These yield an estimate of  a {\em time is money\/} factor, $\mu$, for different transportation scenarios.  In 2010, the value of time for urban travel (all modes) in the Paris region is estimated as 10.7 €/h, with an additional waiting time factor of 1.5, and the guideline that the value of time should increase with GDP {\em per capita\/} with an elasticity of 0.71. We derive $\mu$ for 2022 based on these guidelines as $\mu = 16.65$ €/h. Then we compute the linear conversion from $\tau_e$, to $c_e$, yielding the additive inconvenience cost term, $c_e$, in the objective \eqref{eq:optimal_battery}:
\begin{equation}
    c_e=\mu\tau_e
    \label{eq:c_e}.
\end{equation}

\subsection{Estimation of $c_e$ with Urban Mobility Simulations}
\label{sec:simulation}

In Section \ref{sec:inconvenience_time}, we derived an analytic equation (\ref{eq:expanded_inconvenience_time}) to compute the charging inconvenience time of EVs. 
Such an analytic equation is however based on a number of assumptions, most notably, a uniform density of CSs in the city; a Poisson model of queues; an average speed for all vehicles looking for a CS; and a Euclidean distance to compute charging detour distances. Such assumptions had been introduced to obtain a tractable final equation for charging inconvenience time, however, they may not fully capture the complexities of real-world urban environments.

In principle, the aforementioned quantities could be measured after appropriately sensorizing urban environments. In the following, we shall adopt a mobility simulator for the same purpose, and we shall obtain actual vehicle speeds, trip lengths, traffic impacts, and knowledge of the precise location of CSs within cities. This approach more accurately takes into account the complex interactions between mobility demand, battery capacity, and urban charging infrastructure. The simulation is performed using the Simulation of Urban Mobility (SUMO) software package \cite{behrisch}. By comparing the simulation results with our analytical model, we aim to validate the analytical assumptions and assess their impact on the estimation of $c_e$.

\subsubsection{Simulation Model Overview}
\begin{figure}[ht]
  \centering
  \includegraphics[width=0.6\textwidth]{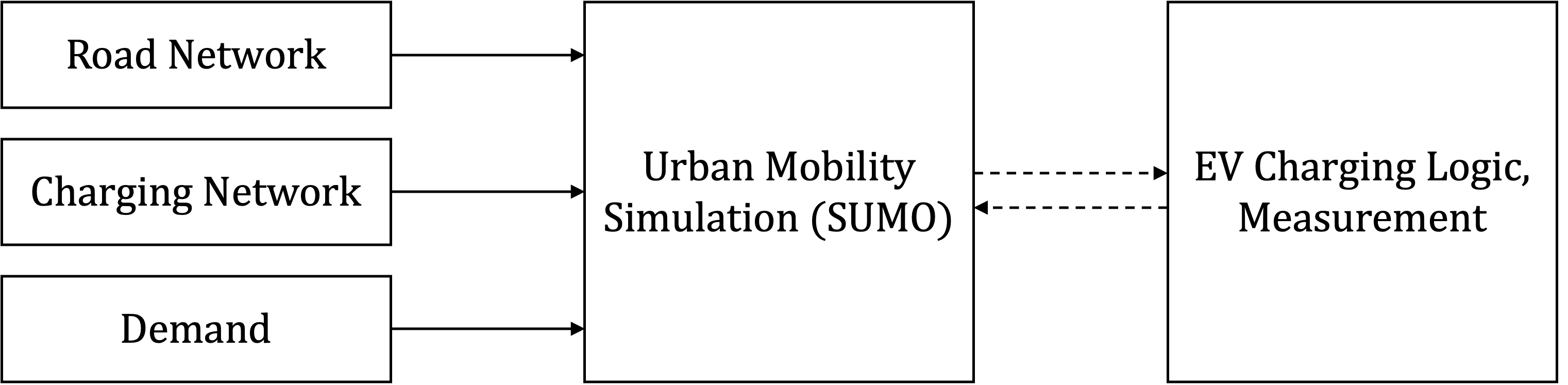}
  \caption{Overview of the main components of the simulation tool used to measure EV charging inconvenience.}
  \label{fig:simulation_overview}
\end{figure}

Our simulation framework consists of five main components, visualised in Figure \ref{fig:simulation_overview}: the road network, the charging infrastructure and mobility demand data are all inputs to SUMO, to recreate city-wide vehicular flows and road traffic. Then a \textit{Python} script interacts dynamically with the simulation, determining when and where EVs charge based on their real time battery SoC (State of Charge); finally, the time spent for the charging processes is collected and measured to estimate the EV inconvenience. 

\paragraph{Road Network:}
We model the urban road network using \textit{SUMO}, which includes roads (edges) and intersections (nodes). Networks are either synthetically generated grids or imported from real-world data sources like \textit{OpenStreetMap} \cite{openstreetmap}. This allows for realistic representation of urban layouts, including varying road lengths, speeds, and traffic controls.

\paragraph{Charging Infrastructure:}
CSs are integrated into the road network with specified locations and charging rates. In our simulations, all CSs have a charge rate of 20 kW, unless specified otherwise. Charging vehicles are parked off-road to prevent congestion.

\paragraph{Mobility Demand:} 
We generate synthetic travel demand using \textit{SUMO}'s \textit{ActivityGen} tool, which follows the activity-based demand modelling approach and computes time-stamped sequences of origin-destination trips for each EV based on demographic data (e.g., working hours, unemployment rate, population and workplace densities, location of schools...). Trips vary in their purpose, duration and start time. For abstract scenarios, demographic data are assigned to resemble urban conditions.

\paragraph{EV specifications:} 
Vehicle parameters influencing energy consumption and performance are specified following sample parameters based on a \textit{KIA SOUL EV 365} that are detailed in the \textit{SUMO} documentation \cite{sumodocumentation}. These parameters include battery capacity, mass, regenerative braking, energy efficiency, wind resistance, energy expenditure for ancillary services (e.g., heating/cooling) and others. Battery capacity B is varied during simulation runs, other parameters are assumed to remain constant.

\paragraph{Charging Logic:}
Each EV follows a stochastic charging logic to decide when and where to charge. First, EVs initiate charging when the State of Charge (SoC) falls below a certain capacity threshold, which is modeled as 
$ C_s \sim \Gamma(\alpha_c = 4, \theta_c = 5).$  
Upon reaching a CS, if a charger is available, the EV begins charging. However, if the station is occupied, the EV may either redirect to the next station with a probability of $P_r = 0.7$, or return to its previous route and defer charging. EVs charge until they reach a desired capacity, modeled as 
$ C_d \sim \Gamma(\alpha_d = 85, \theta_d = 1), $
with a cap at 100\%. 

\paragraph{Data Collection and Inconvenience Estimation:}
During simulations, we record the time each EV spends searching for CSs and actively charging the battery and calculate the total inconvenience time $\tau_e$ for each EV as:
\begin{equation}
    \tau_e = \tau_{\text{search}} + \tau_{\text{charge}}. \label{eq:taue_e_simulation}
\end{equation}
By aggregating $\tau_e$ across all EVs we compute the average charging inconvenience time $\overline{\tau_e}$ and convert into $\overline{c_e}$ using the conversion method specified in Section \ref{sec:conversion}

\section{Simulated E-Mobility Scenarios}
\label{sec:scenarios}

To evaluate the charging inconvenience cost $c_e$ and validate our analytical model, we consider several electric mobility scenarios using both the simulation approach described in Section \ref{sec:simulation}, and the analytical approach presented in Section \ref{sec:inconvenience_time}:
\begin{itemize} 
    \item \textbf{Scenario 1}: Effect of varying battery capacities $B$ on charging inconvenience time $\tau_e$;
    \item \textbf{Scenario 2}: Impact of infrastructure density $\rho$ on $\tau_e$;
    \item \textbf{Scenario 3}: Influence of CS placement on $\tau_e$;
    \item \textbf{Scenario 4}: Real-world case study of Central Paris, analyzing the effect of B on $\tau_e$.
\end{itemize}

Each scenario builds upon the general simulation framework previously outlined and is designed to isolate the impact of specific battery and charging infrastructure parameters on the total charging inconvenience time of EVs. The first three scenarios are computed in a simplified 10x10 lattice road network spanning $400 \text{km}^2$. While this road network has been artificially created for the purpose of this simulation, many modern cities have a similar grid-like structure.
The fourth scenario is computed based on the realistic environment of Central Paris and relies on additional data inputs and processing steps, which are outlined in Section \ref{sec:realistic_scenario}. 
Each simulation run computes realistic mobility behaviour for a simulated time of 30 days. Multiple simulation runs are performed for each scenario, varying an independent variable between runs and measuring the effect on the dependent variable $\tau_e$.

\begin{table}[h!]
\caption{Overview of key simulation variables for E-mobility Scenarios}
\label{tab:scenarios}
\begin{tabular*}{\tblwidth}{@{}LLLLL@{}}

\toprule
\textbf{Variable}      & \textbf{Scenario 1} & \textbf{Scenario 2} & \textbf{Scenario 3} & \textbf{Scenario 4} \\ \midrule
IV                     & $B$       & $n_{cs}$   & $cs_{loc}$ & $B$        \\
DV                     & $\tau_e$  & $\tau_e$  & $\tau_e$  & $\tau_e$   \\ 
Road network           & Grid      & Grid      & Grid      & Paris      \\ 
Area (km$^2$)          & 400       & 400       & 400       & 105        \\ 
$n_{ev}$               & 50        & 100       & 50        & 500        \\ 
$n_{cs}$               & 10        & IV        & 5         & 50         \\ 
\bottomrule
\end{tabular*}

\end{table}

To compare simulation results with the analytical model, we derive the necessary input parameters for (\ref{eq:expanded_inconvenience_time}) directly from the simulations. Specifically, we perform a baseline simulation for each scenario, in which charging events are not considered, such that the measured distance in the simulation relates to the originally intended travel demand, not including charging detours. An overview of the derived parameters for each scenario can be found in Table \ref{tab:input_parameters}.

\begin{table}[h!]
\caption{Input Parameters for analytical model  \eqref{eq:expanded_inconvenience_time} measured from simulations}
\label{tab:input_parameters}
\begin{tabular*}{\tblwidth}{@{}LLLLL@{}}
\textbf{Parameter}       & \textbf{Scenario 1} & \textbf{Scenario 2} & \textbf{Scenario 3} & \textbf{Scenario 4} \\ \midrule
$\eta$ (kWh/km)         & 0.085 & 0.087 & 0.085 & 0.173 \\ 
$v$ (km/h)             & 42.4 & 43.2 & 42.4 & 32.1 \\ 
$d$ (km)               & 1185 & 1176 & 1185 & 570 \\ 
$\sigma$      & 0.7 & 0.7 & 0.7 & 0.7 \\ 
$B$ (kWh)               & IV & [20,40,60] & [20,40,60] & IV \\ 
$\xi$                    & 0.035 & IV & 0.07 & 0.0685 \\ 
$\rho$ (cs/km)           & 0.03 & IV & 0.0125 & 0.476 \\ 
$P$ (kW)                & 20 & 20 & 20 & 20 \\ \bottomrule
\end{tabular*}
\end{table}

\subsection{Scenario 1: Effect of varying battery capacity B on charging inconvenience time $\tau_e$}
\label{sec:scenario 1}

In Scenario 1, we evaluate how varying battery capacities $B$ affect the charging inconvenience time $\tau_e$ for EVs operating in a simplified grid network. We simulate a population of 50 EVs over a period of 30 days, varying the battery capacity from 1 kWh to 100 kWh in 1 kWh increments between simulation runs. All EVs rely solely on public charging infrastructure and there are 10  CSs distributed across the network. The  CSs have a charging rate of 20 kW and are spaced uniformly. Key parameters used in this scenario are provided in Table \ref{tab:scenarios}. The input parameters for the analytical model are derived according to Section \ref{sec:scenarios} and the analytical estimate for $\tau_e$ is computed for the same range of battery capacities using \eqref{eq:expanded_inconvenience_time} for the same period of 30 days. Results from Simulation and Analytical Model are compared in the following paragraph.

Figure \ref{fig:S1,2} (a) shows the charging inconvenience time $\tau_e$ as a function of battery capacity $B$, computed using both analytical and simulation approaches.

We observe that there is a general agreement between the analytical estimate and the one from measured data, however the analytical estimate tends to underestimate inconvenience, due to the general simplifying assumptions underneath its computation. We further observe that the charging inconvenience time decreases as the battery capacity increases. This is because larger batteries require fewer charging events over the simulation period, thus reducing the frequency of extra trips to CSs and associated waiting times. However, the diminishing returns of increasing battery capacity are evident; for instance, increasing the battery capacity from 40 kWh to 100 kWh results in only a 7.1\% decrease in $\tau_e$. These findings suggest that while larger battery capacities can reduce charging inconvenience, the benefits plateau beyond a certain point. Therefore, investing in extremely large batteries may not be cost-effective from an inconvenience reduction perspective.

\begin{figure}[h!]
    \centering
    \begin{tabular}{cc}
        (a) & (b) \\
        \includegraphics[width=0.45\textwidth]{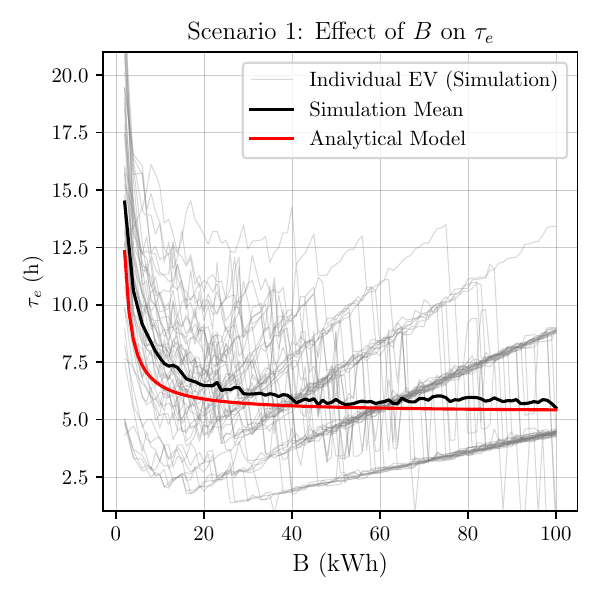} & 
        \includegraphics[width=0.45\textwidth]{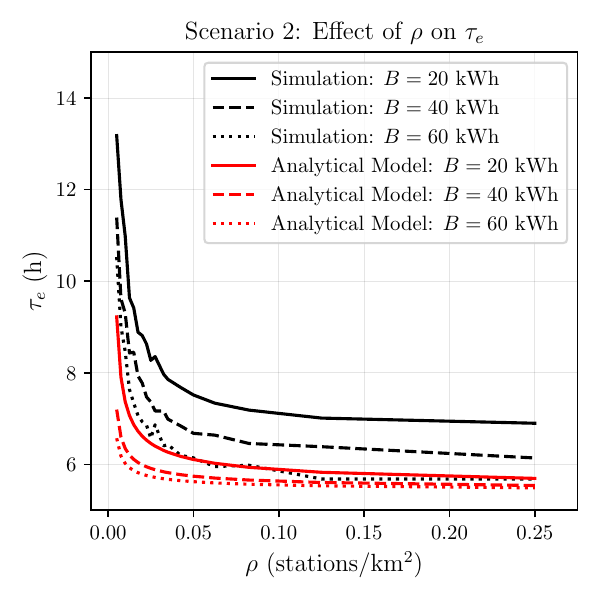} \\
    \end{tabular}
    \caption{(a) and (b) present results for the simulated E-Mobility Scenarios 1 and 2 respectively.}
    \label{fig:S1,2}
\end{figure}

\subsection{Scenario 2: Impact of Infrastructure Density $\rho$ on $\tau_e$} 
\label{sec:scenario2}

In Scenario 2, we explore the impact of infrastructure density $\rho$ on the charging inconvenience time ($\tau_e$). This scenario aims to understand how the density of charging infrastructure affects EV user convenience.

We simulate a population of 50 EVs with a fixed battery capacities of 20, 40, and 60 kWh over a period of 30 days. The number of CSs is varied from 1 to 100 in increments of 1, reflecting an increasing charging station density. The locations of the CSs are randomly assigned across the network. Importantly, when a new CS is added, the positions of existing stations remain unchanged to isolate the effect of increasing $n_{cs}$. Again, key simulation parameters are summarised in Table \ref{tab:scenarios} and the analytical estimation is performed according to the parameters in Table \ref{tab:input_parameters}

Figure \ref{fig:S1,2}(b) illustrates the relationship between the charging inconvenience time $\tau_e$ and the ratio of EVs to CSs. Again, it can be observed that while there is a general agreement between the analytical model and the one based on measured data, the analytical model tends to underestimate the inconvenience.

The results indicate a similar diminishing effect of $\rho$ on $\tau_e$. As the density of CSs increases, the inconvenience time initially decreases significantly, before the additional benefit of higher density diminishes. For example, a battery capacity of 20 kWh may be as convenient as a battery capacity of 40 kWh, if it is possible to increase the charging station density from 0.25 to 0.6 [stations/km$_2$]. This suggests that improving charging infrastructure density can be an effective alternative to increasing battery sizes in reducing charging inconvenience.

\subsection{Scenario 3: Influence of CS Placement on $\tau_e$} 
\label{sec:scenario_3}

\begin{figure}[ht]
  \centering
  \includegraphics[width=0.7\textwidth]{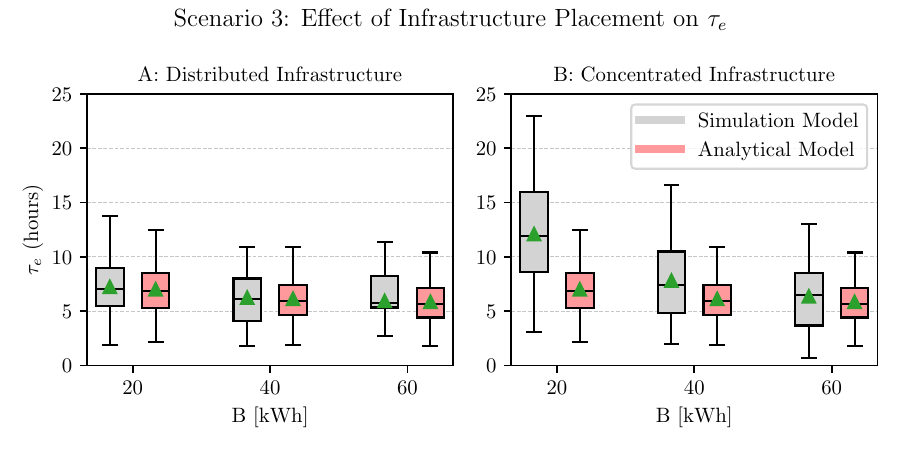}
  \caption{Charging infrastructure placement strategies A (distributed charging infrastructure) and B (concentrated charging infrastructure in a single peripheral point).}
  \label{fig:cs_location}
\end{figure}

In Scenario 3, we investigate how the geographical placement of CSs affects the charging inconvenience time $\tau_e$. Specifically, we compare two different placement strategies to assess the importance of infrastructure distribution. We consider two placements for five CSs:
In Placement A, CSs are uniformly distributed across the road network;
In Placement B, CSs are concentrated along a single road in the top-left corner of the network, as depicted in Figure \ref{fig:cs_location}.
We simulate EVs with battery capacities of 20 kWh, 40 kWh, and 60 kWh over a period of 30 days; additional simulation parameters are summarized in Table \ref{tab:scenarios}. 

Figure \ref{fig:cs_location} compares the charging inconvenience time $\tau_e$ for the two placement strategies across different battery capacities: in Placement A (left), the CSs are randomly placed in the city; in Placement B (right), all CSs are placed in a single peripheral point of the city. In this scenario, it is important to remark that the analytic model can not distinguish the different cases, due to its assumption regarding the position of CSs. However, as can be noticed by the results based on measured data, the placement of CSs does play an important role in decreasing inconvenience, and especially in the case of small batteries, inconvenience is doubled by a poor placement. In other words, by placing CSs in a smart way it is possible to reduce the size of the batteries of EVs without affecting the inconvenience of drivers (the inconvenience of drivers with batteries of $60 kWh$ in Placement B is the same of that of drivers with batteries of $20$ kWh in Placement A.

\section{Realistic Case Study: Central Paris}
\label{sec:central_paris}

We now consider the more realistic urban network corresponding to the centre of Paris (France) to validate the results obtained in the abstract case study described in Section \ref{sec:scenario 1}. We follow the same simulation method outlined in Section \ref{sec:simulation}, and construct a data-driven mobility simulation for Central Paris. We now describe in greater detail the different components and steps that were used to create the case study.

\subsection{Data Sources and Processing steps}

\begin{figure}[h!]
    \centering
    \begin{tabular}{cc}
        (a) & (b) \\
        \includegraphics[width=0.45\textwidth]{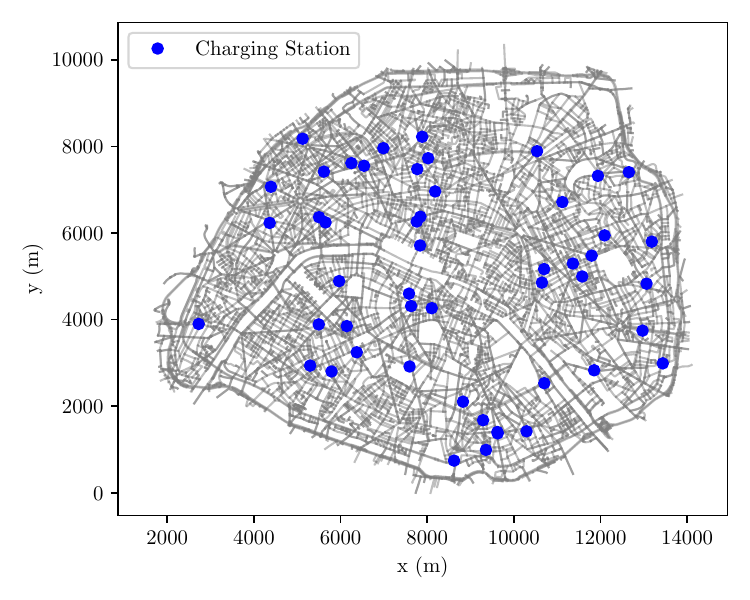} & 
        \includegraphics[width=0.45\textwidth]{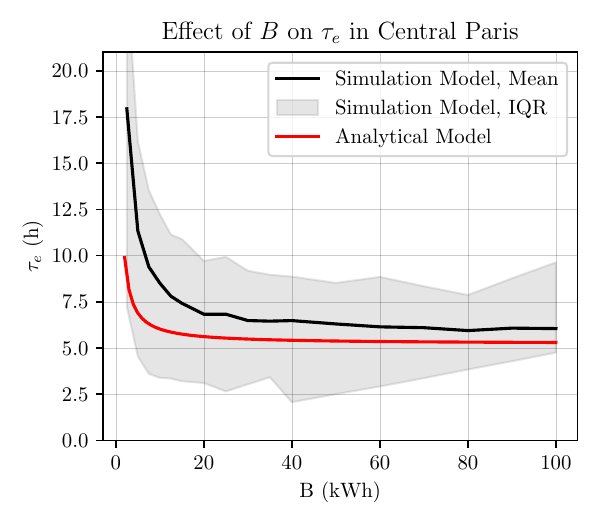} \\
    \end{tabular}
    \caption{(a) illustrates the road network of Central Paris with 50 randomly selected CSs; and (b) presents the results for the simulated Paris Scenario.}
    \label{fig:combined}
\end{figure}

We construct a data-driven mobility simulation for Central Paris using real-world data sources. The simulation setup follows the general framework outlined in Section \ref{sec:simulation}, but includes additional data sources and processing steps:

\paragraph{Road Network:} The road network data for Central Paris is imported from Open Street Map. Post-processing of the converted network involves simplifying junctions and the removal of minor disconnected network regions.
\paragraph{Charging Infrastructure:} Charging infrastructure data for Central Paris are obtained from the National Directory of Charging Infrastructure for EVs in France (IRVE) \cite{data.gouv.fr}. After filtering and removal of duplicates, 381 CSs with a total of 1893 individual plug points have been included. These CSs have been assigned to the SUMO road network of Central Paris based on their geographical coordinates. We have considered all CSs to be equipped with a single plug with charge rate of 20 kW (as before), and since we have considered 500 EVs, we randomly select 50 CSs among the existent ones, to maintain an EV-to-CS ratio of 10. Figure \ref{fig:combined} shows an example of realization of the 50 CSs, in blue dots. 
\paragraph{Mobility Demand} To model synthetic demand for Central Paris, demographic data such as the number of households, car ownership rate, unemployment rate, age distribution, working hour distribution, as well as the locations of schools and population and workplace density across Paris' 20 {\em arrondissements\/} have been obtained from the French \emph{Institut National de la Statistique et des Études Économiques} (INSEE) \cite{insee}. We have processed the data into a statistics file which serves as input to SUMO's \textit{activitygen} tool \cite{sumodocumentationa}. The SUMO  \textit{activitygen} script is used to compute synthetic vehicular mobility demand for 500 vehicles based on these demographic statistics, resembling realistic traffic demand in Central Paris. We have then compared the resulting synthetic demand data with data from the \emph{Mobilité des personnes} survey from the \emph{Ministry of Ecological Transition and Territorial Cohesion} \cite{data.gouv.fra} for validation. In particular, we have found the estimated daily vehicular travel distance, which is a key parameter in our simulations to mimic energy demand of EVs, is consistent with the data indicated in the survey. 

\subsection{Effect of Varying Battery Capacities $B$ on $\tau_e$ in Central Paris (Scenario 4)}
\label{sec:realistic_scenario}
We now present results for a public charging scenario in Central Paris, where a population of 500 EVs is simulated with $B$ (kWh) $= 2.5,\allowbreak 5,\allowbreak 7.5,\allowbreak 10,\allowbreak 12.5,\allowbreak 15,\allowbreak 20,\allowbreak 25,\allowbreak 30,\allowbreak 35,\allowbreak 40,\allowbreak 50,\allowbreak 60,\allowbreak 70,\allowbreak 80,\allowbreak 90,\allowbreak 100$ for a period of 30 days. All EVs rely only on a public charging network consisting of 50 CSs, randomly selected from IRVE data. 

Figure \ref{fig:combined}(b) shows the results of this scenario, and we can observe that the outcome of the abstract case study is replicated for the case of the centre of Paris as well. In particular, the inconvenience initially decreases significantly for greater battery capacities. However the rate of decrease slows significantly for battery capacities greater than 20 kWh. For instance, quintupling the battery size from 20 kWh to 100 kWh enhances the convenience of EVs by only 11.4\%. The analytical model again yields lower estimates for $\tau_e$ compared to the simulation model due to the assumptions introduced in Section  \ref{sec:inconvenience_time}.

\subsection{Optimal Battery Capacity $B^o$ in Central Paris}
\label{sec:results:B^o}

In this section, we now aggregate the introduced notion of inconvenience with the price of a battery, to determine what would be the ``optimal'' size of the battery for drivers in the scenario of Central Paris previously illustrated. 

For this purpose, we convert inconvenience time into a monetary unit according to \eqref{eq:c_e}.
Then, we solve the optimization problem introduced in \eqref{eq:optimal_battery} using both inconvenience cost and purchase cost as derived in Section \ref{sec:c_p}. In this computation we also take into account the ageing of the battery, considering a battery health factor of 0.7, e.g. we estimate the usable battery capacity after 10 years to be 70\% of the advertised battery capacity.

\begin{figure}
    \centering
    \includegraphics[width=0.6\linewidth]{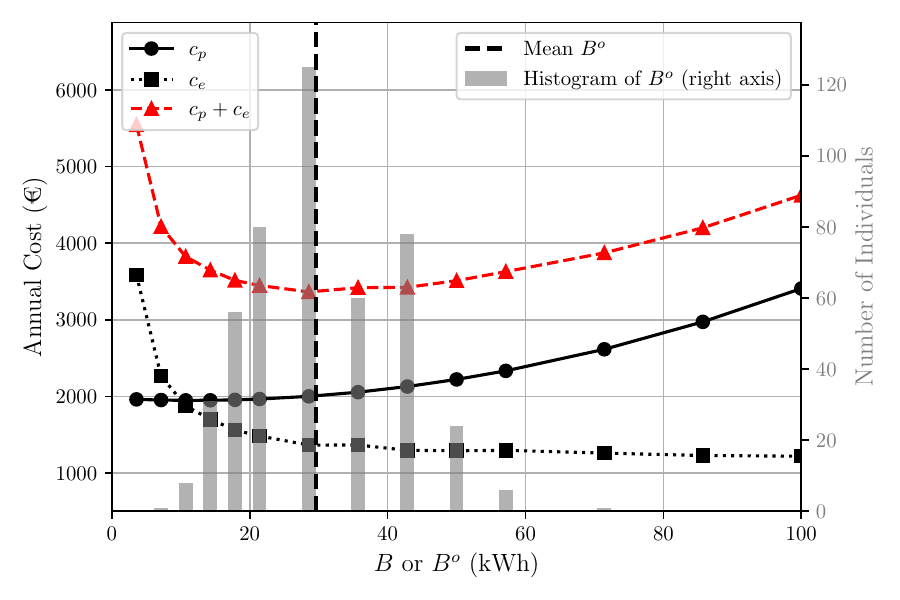}
    \caption{The optimal battery capacity framework introduced in \eqref{eq:optimal_battery}, using data obtained from the Paris simulation.}
    \label{fig:optimal_battery}
\end{figure}

The line graphs in Figure \ref{fig:optimal_battery} illustrate an example of annualized inconvenience cost ($c_e$), EV purchase costs ($c_p$), and their sum (the objective function in \eqref{eq:optimal_battery} for a single driver in our Paris simulation).
We observe that the (composite) objective function is  convex, and we minimise it to obtain the optimal battery capacity for this specific driver based on \eqref{eq:optimal_battery}. Performing this minimisation for all drivers in the Paris simulation leads to a distribution of optimal battery capacities, which is displayed as a grey histogram in Figure \ref{fig:optimal_battery}. We find that the cost-optimal battery capacities for drivers in Central Paris predominantly fall within the range of $15$ - $45$ kWh ($\text{mean} = 29.62 \;\text{kWh}, \;\text{IQR} = 14.28 \;\text{kWh}$). In contrast, the battery capacities of EVs available on the European market in 2023 are significantly larger. Even small cars available on the European market (Classes A and B based on the European passenger car classification), have a mean battery capacity of $47.78$ kWh (IQR = $20.20$ kWh)  \cite{ev-database.org}. This disparity highlights that current market offerings are skewed towards large battery sizes, which diverge from the optimal capacities for urban driving needs revealed in this study. 

\section{Influencing $B^o$ through Infrastructure and Incentives}
\label{sec:influencing_b^o}

We have shown that there is an optimal battery capacity, $B^o$, that minimizes the total cost associated with driving an EV, comprised of vehicle purchase cost $c_p$, and charging inconvenience cost $c_e$. Both $c_p$ and $c_e$ are influenced by the battery capacity $B$, however there are other factors at play, which may be strategically influenced in order to shift $B^o$ towards smaller and more sustainable capacities. This section explores how $B^o$ may be influenced through two primary avenues: charging infrastructure investment; and tax and incentive programmes.

Charging infrastructure investments may aim to increase the density of CSs or upgrade towards higher powered chargers. This influences the charging inconvenience time $c_e$ (see Scenario 2 in Section \ref{sec:scenario2}).
Tax and Incentive policies on the other hand, can affect the effective purchase price ($c_p$) of EVs, for example by penalizing excessive battery capacities and subsidizing smaller, more sustainable ones.

In the following subsection, we compute the effect of charging infrastructure parameters ($P$ and $\rho$) on the objective function \eqref{eq:optimal_battery}, following the same assumptions of the case study in Paris. 

\subsection{Influencing $B^o$ with Investments in Charging Infrastructure}

\begin{figure}[ht!]
    \centering
    \includegraphics[width=1\linewidth]{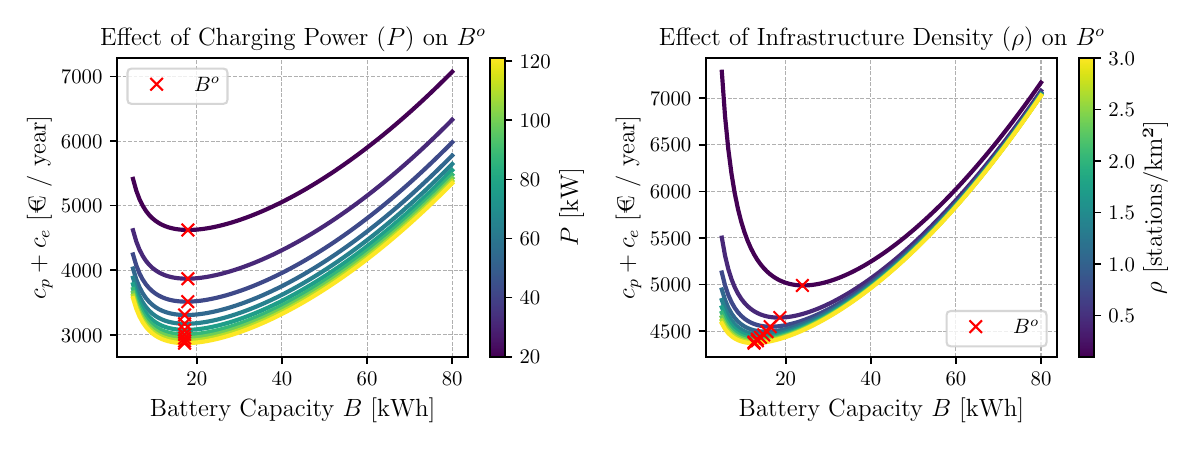}
    \caption{Effect of Charging Power $P$ and Infrastructure Density $\rho$ on Optimal Battery Capacity ($B^o$)}
    \label{fig:influencing_B^o_infrastructure}
\end{figure}

Figure \ref{fig:influencing_B^o_infrastructure} shows the effect of charging power, $P$, and Infrastructure density, $\rho$, on the total cost function \eqref{eq:optimal_battery}. $B^o$ is highlighted in red. 
We can observe that improvements in charging power P, while decreasing inconvenience, do not significantly affect $B^o$, since reductions in $c_e$ are approximately equal for vehicles of different battery capacities. Improvements in Charging Infrastructure density, however, have an outsized effect on $c_e$ for small battery capacities, while larger battery capacities do not benefit as much from higher density infrastructure. Therefore, investments in infrastructure density shift $B^o$ towards smaller capacities.

\subsection{Influencing $B^o$ with Tax and Incentive Policies}
\begin{figure}
     \centering
     \includegraphics[width=0.8\textwidth]{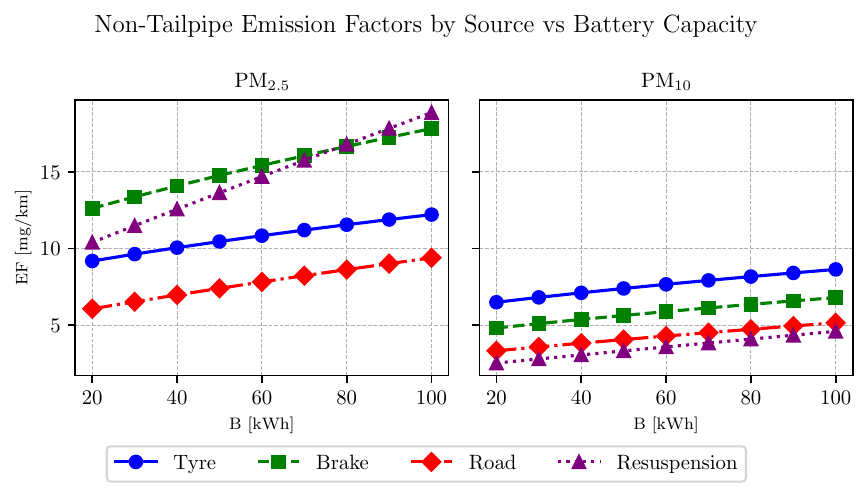}
     \caption{$PM_{2.5}$ and $PM_{10}$ NTE factors as a function of $B$.}
     \label{fig:PM2.5_EF_by_source}
 \end{figure}

Battery-obese cars incur a monotonically increasing environmental and societal toll, for instance, in terms of safety concerns. Accordingly, it may be interesting to tailor appropriate taxes and incentives to modify the utility function (\ref{eq:optimal_battery}), with the ultimate goal of driving EV purchases towards smaller batteries. The easiest way to achieve this objective is to add taxes for battery capacities exceeding, say, 40 kWh, while subsidies may be introduced for capacities below this threshold.
\\
In addition to safety concerns, Non-Tailpipe Emissions (NTEs) is an important instance of environmental externalities. Rather than changing battery (and thus, EV) prices, our methodology could explcitly accommodate such externalities by augmenting the utility function (1) with an additive term, $c_{NTE} (B) > 0$. For instance, \cite{beddows2021} links NTE factors (in mg of  PM/km) to vehicle weight, and assuming again that EV weight is primarily influenced by the battery capacity, a linear relationship may be established between NTE factors and the size of the battery. In particular, using vehicular data of 209 vehicles collected by Weiss et al. \cite{weiss2020}, after removing micro-mobility and heavy goods vehicles, Figure \ref{fig:PM2.5_EF_by_source} depicts such a relationship between battery size and PM2.5 and PM10 NTE factors.
 However, a utility function modified to include the further linear penalization term to address NTE factors would involve adding a {\em socialized\/} environmental impact to the otherwise {\em privatized\/} (i.e.\ driver-specific) utility function (\ref{eq:optimal_battery}).  Furthermore,  the privatized impact of NTE (or, more generally, of  impacts on the environment and on society) would need to be converted to a driver-specific monetary unit (Euro) via an appropriate conversion factor, as was done in Section \ref{sec:conversion} for the inconvenience time, $\tau_e \to c_e$. Therefore, in this paper, we are only opening up a more general discussion of how to include such other costs, particularly  NTE costs, in the objective function for optimisation of  EV battery capacity for individual drivers in their specific urban environment, and subject to specific environmental policy structures. We defer  the full development of this topic to a future work.

\section{Conclusion}
\label{sec:conc}

The trend towards increasing EV battery capacities has significant social, economic, and environmental consequences. In this paper, we introduced a quantitative framework to show that enlarging batteries is not the only way to improve the convenience of EV owners, and that battery enlargement may in fact not be very effective anyway beyond a certain capacity. Extensive simulations---carried out in synthetic grid networks and via a realistic case study of Central Paris---allow us  uniquely to capture the dynamics of two fundamental and ``partly interchangeable'' \cite{wenig2019} components of EV systems: batteries and charging infrastructure. Indeed, investments in charging infrastructure can provide similar convenience benefits for EV users, while at the same time lowering the optimal battery capacity, encouraging the mass adoption of smaller, affordable EVs that lead to less pollution and safer roads.

Our analysis reveals the following key results:
\begin{itemize}
\item Larger batteries decrease EV charging inconvenience, but the benefits of battery size diminish after a certain capacity;
\item Similar improvements in charging inconvenience can be achieved by strategically optimizing charging infrastructure e.g., power, density, and placement;
\item Optimal battery capacities, which minimize the costs associated with both vehicle purchase and charging inconvenience, are significantly smaller than those of current EV models available on the European market;
\item Infrastructure investments and tax/incentive policies can affect charging inconvenience costs and purchase costs respectively,  thereby influencing the optimal battery capacity.
\end{itemize}

Our findings integrate  conflicting perspectives on the trade-off between battery capacity and infrastructure \cite{wenig2019,jones2023a,neubauer2014} into a unified quantitative framework. This framework can be used to identify optimal battery capacities in a given charging environment, but also to estimate the convenience benefits of charging infrastructure investments, providing a nuanced framework for optimizing the design of cost-effective, convenient and sustainable EV systems.

\subsection{Implications for Policy Makers and Manufacturers}

The framework introduced in this paper can inform policies that encourage sustainable EV adoption, specifically by estimating the impact of charging infrastructure investments on EV charging inconvenience and optimal battery capacities.
Automotive companies can leverage these insights to introduce EVs with battery capacities that align with optimal levels in target markets, potentially reducing production costs and environmental impact while meeting consumer convenience needs.

\subsection{Limitations}

While our study provides valuable insights, several limitations should be acknowledged. 
Firstly, the analytical estimation of charging inconvenience relies on assumptions such as a uniform CS distribution, average speeds and a linear conversion from inconvenience time to cost.
Secondly, our simulations, though comprehensive, are based on activity-based demand models and may not fully reflect the variability in individual driving and charging behaviours.
Thirdly, the case study in Section \ref{sec:realistic_scenario} focusses on Central Paris. Although urban centres share common characteristics, results will vary with geographic context.
Finally, there may be other factors that influence EV purchase decisions beyond cost and convenience.

\subsection{Future Research Directions}

\begin{enumerate}
    \item Account for social and environmental externalities in the design of optimal battery capacity: extend the optimization to include social and environmental externalities of battery capacity, such as non-tailpipe pollution and road safety.
    \item Optimal configuration of battery capacity vs infrastructure: existing studies, including our own, do not consider the combined costs of deploying charging infrastructure and vehicles, both of which are required for successful EV systems. Future research could quantify the total costs of EV systems and identify optimal configurations of battery capacity and infrastructure that minimize total system cost while maximising convenience. Such an analysis could reveal important implications of EV system design: Who carries the costs and owns the assets. Small batteries with extensive infrastructure socialize costs and assets, while large batteries with minimal infrastructure privatize them. Future research could investigate the trade-off between batteries and charging infrastructure from this perspective.
    \item Time is money in public charging: In  \eqref{eq:c_e}, we convert charging inconvenience time into cost using a linear conversion factor $\mu$, which represents the valuation of time  during urban travel with a waiting factor as the closest existing proxy. Future research could specifically quantify the value of time during public charging events.
\end{enumerate}

\section{Funding Sources}
This work was supported by the IOTA Foundation; and Google Research.

\section{Declaration of interests}
The authors declare that they have no known competing financial interests or personal relationships that could have appeared to influence the work reported in this paper.

\printcredits

\bibliographystyle{IEEEtran}

\bibliography{BEVI_Paper}



\end{document}